\documentstyle[preprint,aps]{revtex}
 
\begin{document}
\draft
\tighten

\preprint{\vbox{
\hbox{IC/95/145}
\hbox{SISSA Ref. 81/95/A}
}}

\title{
 { IS THERE A MONOPOLE PROBLEM ?}}

\author{Gia  Dvali\thanks{dvali@surya11.cern.ch. Present address:
 CERN, CH-1211 
Geneva 23, Switzerland.}}
\address{ Dipartimento di Fisica, Universita di Pisa
and INFN, Sezione di Pisa I-56100 Pisa, Italy {\rm and}  \newline
Institute of Physics, Georgian Academy of Sciences, 380077 Tbilisi
Georgia}
\author{Alejandra Melfo \thanks{ 
 melfo@stardust.sissa.it.}}
 \address{International School for Advanced Studies,  34014 Trieste, Italy,
 {\rm and} \newline Centro de Astrof\'isica Te\'orica, Universidad de Los 
Andes, M\'erida 5101-A, Venezuela}
\author{Goran Senjanovi\'c \thanks{  goran@ictp.trieste.it}}
\address{International Center for Theoretical Physics
34100 Trieste, Italy }

\maketitle

\begin{abstract}
\noindent We show that there exists a range of parameters in 
$SU(5)$ theory for 
which 
the GUT symmetry remains broken at high temperature, thus avoiding the phase 
transition that gives rise to the overproduction of monopoles. The thermal
production of monopoles can be  naturally suppressed, keeping their number
density below the cosmological limits.  
\end{abstract}

\vspace{2cm} 
 \paragraph*{ A. Introduction}   
 It has been known for a long time that the existence of magnetic monopoles 
(a single one would suffice) would lead to the quantization of
 electromagnetic charge. In grand unified theories based on a simple group
 (or their products), the electromagnetic charge is necessarily quantized
 and thus the magnetic monopoles are the necessary outcome of the theory. 
This, what should be a blessing, is however precisely what makes these 
theories incompatible with  standard cosmology.

Namely, it is believed that at  high temperature in the early universe 
the spontaneously broken grand unified symmetry gets restored. If so, 
during the subsequent  phase transition the monopoles get produced via the 
well-known
 Kibble mechanism \cite{k-77} whenever the original symmetry based on
 a simple group $G$ gets broken down to a subgroup $H$ which contains 
(at least one) $U(1)$ factor. The trouble is that the resulting monopole number
 density $n_M$ would then be some ten orders of magnitude bigger than
 the critical density of the Universe \cite{p-79}.

The crucial assumption in the above is the existence of a phase transition 
that separates the broken and the symmetric phase. The aim of this Letter
 is precisely to address this issue,
namely, to see whether symmetry nonrestoration at high temperature 
\cite{w-74,ms-79} can avoid the monopole problem.

 Previous approaches to the solution of these problem are well known.  One is
of course inflation \cite{g-81}.  Unfortunately, no satisfactory model of
inflation   resulting from a  realistic particle physics
theory exists at present, and in view of this it is of extreme importance to
study alternative possibilities.  Among ``noninflationary'' attempts we want to
cite the one by Langacker and Pi \cite{lp-80} who have argued that a period of
``temporarily'' broken $U(1)_{em}$ in some high temperature interval may avoid
the problem, due to a rapid annihilation of monopoles (produced in a phase
transition at higher $T$) during this period.

In the present paper we want to take a more radical approach and argue that the
phase transition which would produce the monopoles may not take place at all.
The fact that symmetries may remain broken at High $T$ was already noticed
\cite{w-74,ms-79}, and recently \cite{ds-95} it was shown that this effect may
avoid the domain wall problem even in the minimal schemes of physically
important discrete and continuos global symmetries, such as CP or Peccei-Quinn
symmetry.  However, symmetry nonrestoration is not a priory enough to solve the
problem, since unwanted defects can be produced by thermal fluctuations. 
 In the
case of domain walls and global axionic strings, it was shown \cite{ds-95} that
thermal production can be naturally suppressed for a wide range of parameters.
However, there is a crucial difference in the case of monopoles:  domain walls
(or axionic strings) are global defects and can be produced by gauge singlet
fields, therefore there is a rather large choice of parameters for the
suppression of their production rate.  The scenario for monopoles turns out to
be dramatically different and more restrictive, since it is controlled by the
value of the gauge couplings.

The important question for us is whether or not (and under which conditions),
the symmetry gets restored in the minimal realistic GUTs.  Here we analyze the
usual prototype grand unified theory based on the $SU(5)$ gauge group in its
canonical form.  The heavy Higgs field responsible for the $SU(5)$ breaking is
taken to be in the {\bf 24} -dimensional adjoint representation $H_{24}$, and
the light Higgs fields that break the standard model symmetry must belong to
 the
{\bf 5} and {\bf 45} -dimensional representations $\Phi_5$ and $\Psi_{45}$.
  The
minimal model is normally taken to consist of $\Phi_5$ only; whereas the
 minimal
realistic theory of fermion masses is believed to require the existence of
$\Psi_{45}$ too.

What is crucial for the monopole problem is whether or not the vev  of $H_{24}$
 vanishes at high temperatures.  In the minimal model case we find that
 $<H_{24}> \neq 0$ at high T seems to be in conflict with the validity of
 perturbation theory, whereas including $\Psi_{45}$ we find that the symmetry
 nonrestoration is possible for a wide range of the parameters.

Of course, avoiding the phase transition with $SU(5)$ nonrestoration does not
automatically solve the problem. One has to suppose that the field is
``initially'' homogeneously distributed inside a region which  is much larger
than an instant horizon size, although smaller than a comoving scale of
the size of the present horizon.
 This amounts to ask that the so-called horizon
problem be solved by some mechanism (such as  primordial inflation), 
and we leave it to the reader to choose her favorite. We emphasize however
that such a mechanism must be invoked 
in any case for the standard cosmological model to be in agreement with 
observation, and that this requirement is not equivalent to the inflationary 
solution to the monopole problem : whatever the solution is, it does not have 
to be related to the scale of symmetry breaking, as long as it is 
implemented at an earlier time. 

 Even without a phase transition and with uniform initial distribution, 
 at high T monopoles can still be thermally produced in  $e^+ e^- $
 collisions, 
as was studied by Turner \cite{t-82}.   Fortunately,
his analysis shows that for $m_M/T \geq 35$ or so (where $m_M$ is the monopole
mass) the relic number density of monopoles is perfectly compatible with
cosmology.  We have studied the impact of this constraint on the broken $SU(5)$
theory at high temperature and our analysis puts the minimal model in serious
trouble, whereas once again the more realistic version with $\Psi_{45}$ works
out right.

 Thus, our work seems to indicate that the monopole problem is not an
 inevitable consequence of grand unification, but rather a dynamical 
question which depends on the  spectrum and the parameters of the theory.

 \paragraph*{ B. $SU(5)$ theory at low and high T}  
\hspace{0.5cm} {\bf a:}
We first study the high T behavior of the minimal $SU(5)$ theory with $H_{24}$
and $\Phi_5$ Higgs fields (we drop their subscripts hereafter).  At $T=0$ the
Higgs potential is

\begin{eqnarray}
V &=& - m_H^2 Tr H^2 + \lambda_1 (Tr H^2)^2 + \lambda_2 Tr H^4 \nonumber \\
&-& m_\Phi^2 \Phi^\dagger \Phi + \lambda_\Phi (\Phi^\dagger \Phi)^2 - \alpha 
\Phi^\dagger \Phi Tr H^2 - \beta \Phi^\dagger H^2  \Phi
\label{pot0}
\end{eqnarray}

where $ H = \sum_{a=1}^{24} H_a \lambda_a$, and $T_a = \lambda_a/2$ are the 
generators of $SU(5)$ for the {\bf 5}
dimensional representation such as $\Phi$.  The desired symmetry breaking
 $SU(5)
\stackrel{<H>}{\longrightarrow} SU(3)_C\times SU(2)_L \times U(1)_Y$ with
$ <H> = v_H \, {\rm diag}(1,1,1,-3/2,-3/2)$ implies the conditions

\begin{equation}
\lambda_2 >0 \;\; , \;\;\;\; 30 \lambda_1 + 7 \lambda_2 > 0\;\; ; \;\;\; 
\beta >0
\label{cond1}
\end{equation}

When the final stage of symmetry breaking is turned on through $<\Phi^T> = 
(0,\, 0,\, 0,\, 0,\, v_\Phi)$, the minimum conditions require further

\begin{equation}
\lambda_\Phi >0 \;\; , \;\;\; (30\lambda_1 + 7 \lambda_2)(40\lambda_2
\lambda_\Phi - {9 \over 2} \beta^2) - 3(10\alpha + 3\beta)^2 >0
\label{cond2}
\end{equation}

The conditions (\ref{cond1}) and (\ref{cond2}) play a crucial role in the
 study of the $SU(5)$ phase diagram at high T. The computation of the
 effective Higgs potential  at high T is rather complicated, but our task
 is facilitated by focusing on the leading terms of order $T^2$. Namely,
 we are interested in the high T phase diagram of $SU(5)$ for $T\gg m_H$,
 and then we need the form of the $T^2$-dependent mass terms for the $H$
 and $\Phi$ fields. 

  In the approximation of weak couplings, assuming the
 validity of perturbation theory one can use  the general expression given by 
Weinberg \cite{w-74}

 \begin{equation}
 \Delta V(T) = {T^2 \over 24} \left[ \left({\partial^2 V \over \partial 
\varphi_i \partial\varphi^i} \right)  +
 3 (T_a T_a)_{ij} \, \varphi^i \varphi^j \right] 
 \end{equation}

where $T_a$ are the group generators and  $\varphi_i$ are the real 
components of the fields. For our potential this gives

\begin{eqnarray}
\Delta V(T) &=& {T^2 \over 24} \left\{ ( 48 \lambda_\Phi - 96 \alpha - 
{96\over 5}\beta + {36 \over 5} g^2) \Phi^\dagger\Phi \right. \nonumber \\
&+& \left. (208 \lambda_1 + {376 \over 5} \lambda_2 - 20 \alpha -
 4 \beta +{15 \over 2} g^2) Tr H^2 \right\} \nonumber \\
&& \nonumber \\
&\equiv&   m^2_\Phi(T) \Phi^\dagger\Phi +
 m^2_H(T) Tr H^2  
\label{potT}
\end{eqnarray}

The above form has already been given in Ref \cite{kst-90}.  Now, since
 $\beta >0$ and $\alpha$ too is allowed to be positive, one cannot make any a
 priori statements about the signs of the mass terms above.  Actually, it was
 already noticed \cite{kst-90} that (\ref{potT}) allows for a negative mass
 for $\Phi$, thus enabling the non-restoration of the $SU(2)_L\times U(1)$
 symmetry.  Since this is achieved at the expense of $\alpha$, $\beta$ being
  positive, it is easily seen that the coefficients in (\ref{potT}) make the
 nonrestoration of $H$ much harder to achieve.

Notice first that the conditions (\ref{cond1}) and (\ref{cond2}) cannot allow
 both mass terms in (\ref{potT}) negative; but what about the coefficient of
 $H$?    It turns out   that the nonrestoration of $<H>$ seems to require
 $\lambda_\Phi >1$ and thus invalidates the weak-coupling expression
 (\ref{potT}).  To see what is going on let us look at the simplified problem
 with $\lambda_2$ and $\beta$ small.  The conditions (\ref{cond1}) and
 (\ref{cond2}) now read ($\lambda_H = \lambda_1$)

\begin{equation}
\lambda_H > 0\; , \;\; \lambda_\Phi > 0\; , \;\; 4\lambda_H \lambda_\Phi > 
\alpha^2
\label{cond3}
\end{equation}

and $m^2_H(T) < 0$ requires  

\begin{equation}
\alpha > {52 \over 5} \lambda_H + {3 \over 8} g^2
\label{cond4}
\end{equation}

It is easy to see that (\ref{cond3}) and (\ref{cond4}) imply

\begin{equation}
\lambda_\Phi > \left({26 \over 5}\lambda_H + {3 \over 16}g^2 \right)^2 
{1 \over \lambda_H}
\label{la5}
\end{equation}

and $\lambda_\Phi$ as a function of $\lambda_H$ has a minimum at $ \lambda_H = 
{15 \over 416} g^2 $. Thus we  have a lower limit for $\lambda_\Phi$

\begin{equation}
\lambda_\Phi \geq {39 \over 10} g^2
\label{glower5}
\end{equation}

Taking a typical value $g^2/(4\pi) \simeq 1/50$, this means 
$\lambda_\Phi \geq 1 $.
 Clearly, the weak coupling limit of (\ref{potT}) ceases to be justified.

Of course, the full computation must include the couplings $\alpha$ 
and $\beta$,
and this requires a numerical analysis.  We have performed it, and the end
result is that (\ref{glower5}) is not modified much.  The point is that the
couplings $\lambda_1$, $\lambda_\Phi$ and $\alpha$ enter with the largest
coefficients in (\ref{potT}), and thus it is more or less their role to
determine whether or not the $SU(5)$ symmetry may remain broken at high T ($T
\gg m_H$)

{\bf b:}  We have seen above that the
 requirement of the validity of the perturbation theory points towards the
 usual
 assumption of the restoration of the $SU(5)$ symmetry.  Now, the analysis was
 performed for the minimal $SU(5)$ model with the light Higgs $\Phi$ being {\bf
 5}-dimensional.  But the minimal theory suffers from the problem of the
 fermionic spectrum being non realistic, namely whereas $m_b \simeq m_\tau$ can
 be considered a success, this  relation fails badly for the first two
 generations.  It is generally believed that the realistic $SU(5)$ theory must
 contain at least a {\bf 45}- dimensional multiplet ($\Psi$) needed to cure
 this
 problem.  This prompted us to perform the above analysis for this, what should
 be considered a minimal realistic theory.  Now, from the expression for the
 high T mass term in (\ref{potT}), it is clear (as we already remarked) that it
 is easier to keep the vev of the smaller representation nonrestored, since
 $\alpha$ enters in its mass term with a much larger coefficient.

The analysis with {\bf 45} parallels the one performed above, and of course it
gets even more messy.  For the sake of space and since it worked well above, we
present the computation in the limit of $\lambda_2$ and $\beta$ small (and the
analogous couplings for $\Psi_{45}$ also small), i.e.  we keep only $\alpha$,
$\lambda_H$ and $\lambda_\Psi$ with $\lambda_\Psi$ defined as in (\ref{pot0}).
More precisely, if we decompose $\Psi$ into 90 real (45 complex) fields
$\Psi_i$, we can write $V(H,\Psi)$ as in (\ref{pot0}) with $\Phi^\dagger \Phi$
substituted by $\sum_{i=1}^{90} \Psi_i^2$.

Again, from the  general form in \cite{w-74}, one can easily deduce the mass
 terms for $\Psi$ and $H$ at high T

\begin{eqnarray}
m^2_\Psi(T) &=& \left( 368 \lambda_\Psi - 96 \alpha + {96 \over 5} g^2
\right) {T^2 \over 24} \nonumber \\
m^2_H(T) &=& \left( 208 \lambda_H - 180 \alpha + {15 \over 2} g^2
\right) {T^2 \over 24}
\label{mass45}
 \end{eqnarray}

Our point about the dimension of the representation and the nonrestoration of
 its vev is manifest in (\ref{mass45}):  
it is clearly much easier to keep $<H>$
 nonzero at high T (than $<\Psi>$).  With the condition for the boundedness of
 the potential

\begin{equation}
\lambda_H > 0\; , \;\; \lambda_\Psi > 0\; ,
 \;\; 4\lambda_H \lambda_\Psi > \alpha^2
\label{cond5}
\end{equation}

we now obtain (instead of (\ref{la5})

\begin{equation}
\lambda_\Psi > \left({26 \over 5}\lambda_H + {3 \over 16}g^2 \right)^2
 {1 \over 81 \lambda_H}
\label{la45}
\end{equation}

Thus we get (instead of (\ref{glower5}))

\begin{equation}
\lambda_\Psi \geq {13 \over 270} g^2
\label{glower45}
\end{equation}

Clearly $\lambda_\Psi$ is allowed to remain small, while having $<H> \neq 0$
at $T>m_H$.

Switching on other couplings in the potential does not change the results
drastically.  The bottom line is that $SU(5)$ may remain broken at high T, thus
avoiding the phase transition which leads to the disastrous overproduction of
monopoles.

\paragraph*{ C. The monopole density}  
As we mentioned in {\it A}, the nonrestoration of symmetry, although 
necessary, is not sufficient to guarantee the non overabundance of monopoles.
 Monopoles  can be thermally produced in $e^+ e^-$   (and other
 charged particles) collisions, and from the analysis by Turner 
\cite{t-82} we know that their density depends crucially on $m_M/T$ at
 these high temperatures. He finds out that in order to be consistent
 with cosmology, we need

\begin{equation}
{m_M \over T} \geq 35
\label{lim}
\end{equation}

More precisely, for $m_M/T \geq 20$, he finds out

\begin{equation}
{n_M \over n_\gamma} \simeq 3\times 10^3 \left( m_H \over T \right)^3 
e^{- 2 m_M/T}
\label{elim}
\end{equation}

where $n_\gamma$ is the photon density; and from the upper limit
 $n_M/ n_\gamma\leq 10^{-24}$, one obtains (\ref{lim})

Now, in $SU(5)$  the lightest monopoles weigh \cite{dt-80}
 
\begin{equation}
m_M = {10 \pi \over \sqrt{2} g} v_H
\end{equation}

For $g^2/ (4 \pi) \simeq 1/50$ or $g \simeq 1/2$, $m_M \simeq 40 v_H$,
 and thus the consistency with the cosmological bound (\ref{lim}) implies

\begin{equation}
{v_H \over T} \geq 1
\label{vt}
\end{equation}

From (\ref{pot0}) and (\ref{potT}), we get for $T\gg m_H$

\begin{equation}
{v_H^2 \over T^2} = - { 208 \lambda_1 + {376 \over 5} \lambda_2 - 20 \alpha -
 4 \beta +{15 \over 2} g^2  \over 12 (30 \lambda_1 + 7 \lambda_2)}
\label{vt2}
\end{equation}
 
Obviously (\ref{vt}) and (\ref{vt2}) will put even more restrictive
 conditions on the parameters of the theory (than just (\ref{glower5}) or
 (\ref{glower45}) ). In any case, the analysis is straightforward and we
 quote the results for the simplified models with only  $\lambda_\Phi$
 ($\lambda_\Psi$), $\lambda_H$ and $\alpha$ couplings in the Higgs
 potential (\ref{pot0}).
 
{\bf a.}   Let us see first what happens for the minimal model with
 $\Phi_5$. For $\lambda_1 = \lambda_H$, from (\ref{cond3}),
 (\ref{vt})
 and (\ref{vt2}) we get

\begin{mathletters}
\label{5cond}
\begin{equation}
\alpha > {142 \over 5} \lambda_H + {3 \over 8} g^2  
\label{valpha5}
\end{equation}
 \begin{equation}
\lambda_\Phi > {213 \over 20} g^2
\label{lav5}
\end{equation}
\end{mathletters}

For $g^2 \simeq 1/4$, $\lambda_\Phi \geq 2.7$ and the perturbation theory
 clearly fails.

{ \bf  b.}
 We repeat the same for the more realistic version with the
$\Psi_{45}$ representation. As before (compare with (\ref{glower5}) and
 (\ref{glower45})), the condition (\ref{lav5}) relaxes by a factor of $1/81$, 
 and we get

\begin{equation}
\lambda_\Psi > {213 \over 1620} g^2
\label{lav45}
\end{equation}

which for $g^2 \simeq 1/4$ would give $\lambda_\Psi > 1/30. $
 Thus, the largest coupling of the theory $\lambda_\Psi$ is still quite
 small and the perturbation theory is operative.

In summary, whereas in the minimal model, at least in perturbation theory,
 the monopole problem persists, in the more realistic version we see that
 it may not be there. Since the realistic theory requires the existence of
 {\em both} $\Phi_5$ and $\Psi_{45}$, the nonrestoration of $<H>$ and the non 
overabundance of monopoles produced  becomes
 only easier to achieve.  

Unfortunately, from   the exponential nature of the monopole density in
 (\ref{elim}), it is clear that due to the uncertainty in the Higgs 
couplings we cannot predict precisely the monopole density.

\paragraph*{D. Summary and outlook}  

Our results seem to indicate that the problem of monopoles may not be generic
 to GUTs. Whether or not there is an overabundance of monopoles is directly
  tied up to whether the GUT symmetry is restored or not, and our analysis
   shows that the symmetry nonrestoration is in general allowed, but it 
   depends on the spectrum and the couplings of Higgs scalars. 

We have studied this issue in the prototype theory of all GUTs, the $SU(5)$
 model, and found out that the problem persists in its minimal version with
 the {\bf 5}-dimensional light Higgs, but that the more realistic variant
 with a {\bf 45}-dimensional Higgs included eliminates (potentially) the 
problem.

We wish to say a few words about the generality and the meaning of our results

\begin{description}

 \item[{\it i})] Unlike inflation, the symmetry non-restoration scenario does
not result in a negligible present-day value of the monopole number density.
Thus, the possibility remains open for monopoles to be 
the required dark matter.
Whether or not the monopole density is large enough to allow for experimental
detection is again related to the spectrum of the theory.

\item [{\it ii})] The important question is what happens in the 
supersymmetric
version of the theory, which is favored from the point of view of the hierarchy
problem and the unification of couplings.  Unfortunately, at the level of the
leading $T^2$ analysis for small gauge couplings,
 it has been shown \cite{mm-84}
(in the context of global supersymmetry) that internal symmetries get restored
at high T.  This would imply the existence of the monopole problem 
in SUSY GUTs.
It is worth investigating, though, the generality of these results, with for
example the non-leading ``daisy'' diagrams contributions to the high T behavior
of the theory, but this is beyond the scope of this paper.

\item [{\it ii})] What about other GUTs, such as $SO(10), E(6)$...? It should
 be clear from our discussion that the situation will depend on the Higgs 
spectrum
 of the theory.  In many popular models one assumes the existence of a large 
representation, such as say {\bf 126} in $SO(10)$, used to provide
 the mass for 
the right handed neutrino. Obviously,  the presence of such a large number of 
fields will help the nonrestoration of the GUT symmetry. 
 We leave the analysis 
of the extended theories (with more
 detail on the high T analysis) for a longer paper in preparation.

\end{description}

 We are grateful to   G. Bimonte, G. Lozano and M. Quir\'os  
 for enlightening discussions, and to R. Brandenberger for important
remarks.   Special thanks are due to C. Aulakh for insightful comments.

 {\bf Added Note:}  
After this paper was accepted for publication, we learned that a similar idea 
has been put forward (and completely ignored in the literature) 10 years
ago by Salomonson,  Skagerstan and Stern \cite{sss-85}.  These 
authors however ignored the effect of the gauge coupling, which as is clear 
from 
our analysis plays an important role. The recent work of Bimonte and Lozano 
\cite{bl-95}, in which they compute the next-to-leading order corrections to 
the effective potential for the models considered here, indicates that 
symmetry non-restoration  may require either a smaller 
gauge coupling, or a complete analysis of the 
general Higgs potential including all the couplings (this work is now in 
progress).

\end{document}